\documentclass[aps,prl,twocolumn,showpacs,superscriptaddress,floatfix]{revtex4-1}
\usepackage{graphicx}
\usepackage{bm}
\usepackage{amssymb}
\usepackage{amsmath}

\begin{document}

\title{Skipping and snake orbits of electrons: singularities and catastrophes}

\author{Nathan Davies}

\affiliation{Physics Department, Lancaster University, Lancaster, LA1 4YB, UK}

\author{Aavishkar A. Patel}

\affiliation{Physics Department, Lancaster University, Lancaster, LA1 4YB, UK}

\affiliation{Department of Physics, Indian Institute of Technology Kanpur, Kanpur 208016, India}

\author{Alberto Cortijo}

\affiliation{Physics Department, Lancaster University, Lancaster, LA1 4YB, UK}

\affiliation{Departamento de Física Teórica, Universidad Autónoma de Madrid, 28049 Madrid, Spain}

\author{Vadim Cheianov}

\affiliation{Physics Department, Lancaster University, Lancaster, LA1 4YB, UK}

\author{Francisco Guinea}

\affiliation{Instituto de Ciencia de Materiales de Madrid, CSIC, 28049 Madrid, Spain}

\author{Vladimir I. Fal'ko}

\affiliation{Physics Department, Lancaster University, Lancaster, LA1 4YB, UK}

\begin{abstract}
Near the sample edge, or a sharp magnetic field step the drift of 
two-dimensional electrons in a magnetic field has the form of 
skipping/snake orbits. We show that families of skipping/snake 
orbits of electrons injected at one point inside a 2D metal 
generically exhibit caustics folds, cusps and cusp triplets, 
and, in one extreme case, a section of the batterfly bifurcation.
Periodic appearance of singularities along the $\pm B$-interface 
leads to the magneto-oscillations of nonlocal conductance 
in multi-terminal electronic devices.
\end{abstract}


\maketitle

Skiping orbits have been introduced into physics of metals by Niels Bohr 
in the early studies of diamagnetism \cite{Bohr-1911}. Skipping orbits 
also play a special role in the two-dimensional (2D) electron systems, 
where they determine the chiral current-carrying properties of the electron 
edge states \cite{Teller} important for the formation of the quantum 
Hall effect \cite{Halperin}. Snake orbits have been discussed, first, 
in the context of the electrons propagation along domain walls 
in ferromagnetic metals \cite{Mints,ShekhterRozhavski} and, later, used 
for channeling ballistic electrons in 2D semiconductors in a spatially 
alternating magnetic field 
\cite{Snakes-1,Snakes-2,Snakes-3,magn-step-1,magn-step-2,magn-step-3}, 
$\mathbf{B}=\mathbf{l}_{z}B\mathrm{sign}x$. 
In isotropic 2D metals, where all electrons with
energies close to the Fermi level revolve along cyclotron circles with the
same radius, $R=p_{F}/eB$, skipping and snake orbits have the form of
consecutive circular segments matched by the specular reflection
at the edge, or by the smooth continuity on the oposite sides of the 
$\pm B$-interface, see Fig. 1. A close relation between these two families 
can be established  \cite{Taddei} by folding a sheet of a 2D material 
(graphene \cite{MLG} or hexagonal transition-metal dichalcogenide 
\cite{GeimPNAS,Coleman} which can sustain sharp bends with the radia much 
smaller than the electron cyclotron radius) in a homogeneous magnetic 
field. Then, the electron path near the 
fold looks like a skipping orbit with circular segments alternating between 
the top and bottom layers, but when projected onto an unfolded sheet, 
the electron motion resembles the motion near a $\pm B$-interface.

\begin{figure}[h]
\includegraphics[width=0.9\columnwidth]{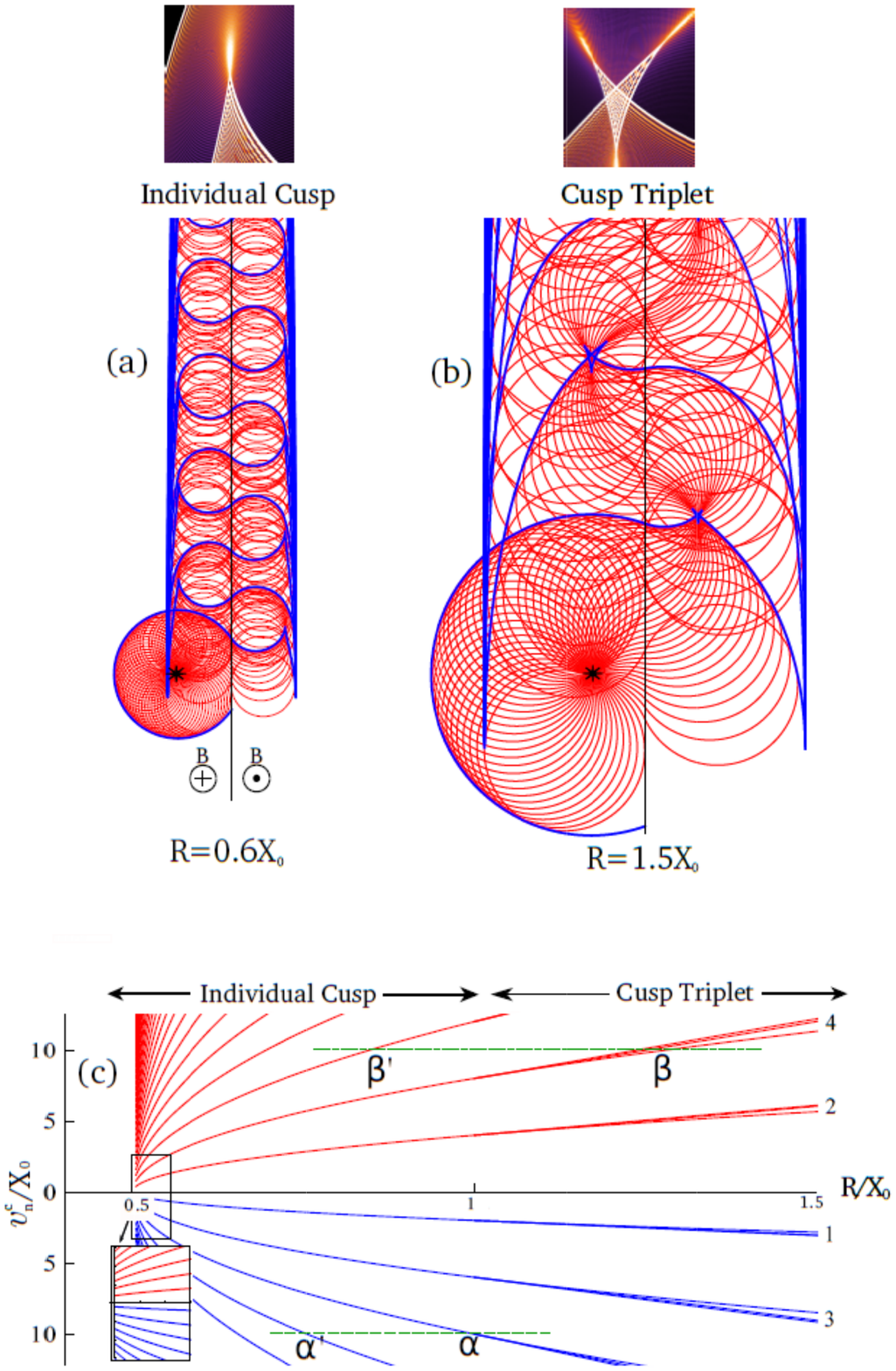}
\caption{Caustics (blue) of snake orbits (red) for: (a) $R<X_0$; (b) $R>X_0$. 
(c) Positions, $v_n^c$ of cusps on the r.h.s (bottom) and l.h.s. 
(top) from the $\pm B$-interface. Top panels show semicalssaically calculated
interference patterns in the vicinity of classical singularities.}
\label{Fig1}
\end{figure}

Below, we study bunching in the families of skipping and snake orbits of
electrons injected into a 2D metal and singularites in the
spatial distribution of electronic trajectories. Mathematically, 
such features originate from the singularities in the differentiable 
maps in Thom's catastrophe theory \cite{Thom,Arnold}.
The list of stable singularities in 2D dynamical systems includes caustic 
folds (caustics) and cusps, whereas focus is the most famous unstable 
singularity. All of those are often encountered in ray optics \cite{footnote}. 
While focusing of light is one of the widest-implemented physical 
phenomena, one often encounters optical caustics when observing sunlight 
sparkling on the sea, or starlight twinkling \cite{Berry}. 
In electronics, observations of caustics and focusing are less common. 
In 2D systems, electron focusing requires designing potential lenses 
\cite{ElectronicLens} and electrostatic mirrors 
\cite{ElectronicInverseLens} by taylor-gating semiconductor 
heterostructures. Caustics and focusing of surface-band electrons have 
also been observed in 'corals' built by the STM manipulation of atoms on 
noble metal surfaces \cite{Eigler}. Focusing has also been realised using 
a mmagnetic field. Magnetic focusing of electrons in the 
three-dimensional space has facilitated studies of Fermi surfaces of pure 
metals and semimetals \cite{Sharvin,Heil,Weismann}.  Later, this magnetic 
focusing technique has lent its name to the nonlocal magnetotransport 
effect observed near the edge of a 2D electron gas in semiconductors 
\cite{vanHouten,Beenakker,Goldman,Smet,Westerwelt}, where, 
inspirationally for this work, one family of caustics has been identified 
\cite{Beenakker} for skipping orbits of electrons injected 
from the edge of a ballistic 2D electron system. 
Here, we demonstrate that caustic bunching is generic for snake/skipping
orbits of electrons injected into a 2D metal at any distance $X_{0}<2R$
near the $\pm B$-interface/sample edge. As shown in Fig. 1, 
for $X_{0}>R>\frac{1}{2}X_{0}$, the networks of caustics displays 
a periodic appearence of individual cusps, which split into cusp triplets 
when $R>X_{0}$. In general, such crossover would happen via the formation 
of swallowtail singularities \cite{footnote}, but, uniquely for a sharp 
field step/sample edge, these two regimes are separated at $X_0=R$ by
the butterfly bifurcation \cite{footnote} seen as a higher-intensity 
unstable singularity of the intensity.

For electrons isotropically injected into a 2D metal with an isotropic 
dispersion of electrons, at a point $(-X_{0},0)$ near the 
$\pm B$-interface/edge at $x=0$, their trajectories can be parametrised using 
the angle $\theta$ (counted in the anticlockwise direction) between the 
initial velocities and $x$-axis. In Figure 1, these trajectories are drawn 
for $0<\theta <2\pi$, with a step of $0.1$. For the orbits near the 
$\pm B$-interface/edge, these are the sequences of
semi-circles with the coordinates $\mathbf{r}_{n}=(x_{n},\ y_{n})$,
\begin{eqnarray}
x_{n} &=&\zeta _{n}+R\sin \varphi ,\;\zeta _{n}=b^{n}(R\sin \theta -X_{0}),
\label{defs} \notag \\
\ y_{n} &=&\eta _{n}+R\cos \varphi ,\;
\eta _{n} = 2n\sqrt{R^{2}-\zeta _{0}^{2}}-R\cos \theta ,  \notag
\end{eqnarray}
where $n=0,1,2,...$ labels the number of times the trajectory arrived at the
interface/edge at $x=0$, angle $\varphi $ counted from the $y$-axis 
allows to describe all points on a single segment 
($b^{n}R\sin \varphi <-\zeta _{0}$, for $n>0$), and $b=\pm 1$ 
for skipping/snake orbits. A sheet density, 
\begin{eqnarray}
\rho = \int \delta (\mathbf{r}-\mathbf{r}_{n})d\varphi d\theta,  \notag
\end{eqnarray}
of such trajectories can be evaluated (using a sequence of substitutions), as 
\begin{eqnarray}
\rho (\mathbf{r})
=\left. [1- b^n \mathrm{sign} x] \left\vert 
\frac{\partial F}{\partial \theta }\right\vert
^{-1}\right\vert _{F=0}, \\
F(x,y;\theta )\equiv (x-\zeta _{n})^{2}+(y-\eta _{n})^{2}-R^{2}.  \notag
\end{eqnarray}
This density is singular along caustics $\mathbf{R}_{n}=(u_{n},v_{n})$, 
where, simultaneously, 
\begin{eqnarray}
\frac{\partial F}{\partial \theta }=0; F=0.
\label{one}
\end{eqnarray} 
This determines the equations for causitcs,
\begin{eqnarray}
u_{n} &=&\zeta _{n}\pm R\frac{c_{n}s}{\sqrt{1+c_{n}^{2}}},\;
c_{n}=\frac{2n\zeta _{0}}{\sqrt{R^{2}-\zeta _{0}^{2}}}-\tan \theta ,  
\notag \\
v_{n} &=&\eta _{n}\pm R\frac{b^{n}s}{\sqrt{1+c_{n}^{2}}},\;
s=\mathrm{sign}(\cos \theta ),  
\label{caustics}
\end{eqnarray}
where we choose the sign '$\pm $' and permitted range
of $\theta $ using the requirement that 
\begin{eqnarray}
\zeta _{0}\sqrt{1+c_{n}^{2}}\pm b^{n}c_{n}sR<0.  \notag
\end{eqnarray}

The density of trajectories is most singular in the vicinity of cusps
($u_n^c,v_n^c$), which are characterised by the condition 
$\frac{d^2 F}{d\theta^2}=0$, additional to Eq. (\ref{one}). 
For strong magnetic fields, 
such that $X_{0}>R>\frac{1}{2}X_{0}$, the periodically appearing 
cusps are illustrated in Fig. 1(a), with $u_n^c=X_0$ and 
the universal curving of caustics near the cusps, 
\begin{eqnarray}
u_n-u_n^c \propto \pm (v_n^c-v_n)^{3/2}.
\end{eqnarray}
A very typical for cusps calculated semiclassically interference pattern is 
shown in the top inset. The y-positions 
of the cusps, $v_n^c$ are plotted in Fig. 1(c) against the ratio $R/X_{0}$. 
The top/bottom parts of Fig. 1(c) distinguish between the cusps appearing on 
the left/right from the magnetic field step. For skipping orbits, 
the latter should be folded onto the same half-plane. When $R<\frac{1}{2}X_{0}$, 
all the extended caustics disappear, leaving only one limiting caustic of the 
closed orbits: a circle with a $2R$-radius centred at the source. The inset in 
Fig. 1(c) shows the limiting positions of the cusps at 
$R\rightarrow \frac{1}{2}X_{0}$ (spaced with the period of $\approx 0.4X_{0}$) 
before their final disappearence. 

The second regime in the formation of catastrophes of snake/skipping orbits is 
characteristic for the weak magnetic fields, such that $R>X_{0}$. 
In the latter case, the singularities form cusp triplets shown in Fig. 1(b). 
The semiclassically calculated interference pattern of the electron waves in 
the vicinity of the cup triplets is illustrated in the top inset in Fig. 1(b), 
and their positions are shown on the r.h.s. of Fig. 1(c). Note that for 
$X_{0}/R\longrightarrow 0$ caustics in Fig. 3(b) transform into 
caustics of skipping orbits originated from a point source 
exactly at the edge of a 2D system \cite{Beenakker}. 

\begin{figure}[h]
\includegraphics[width=0.7\columnwidth]{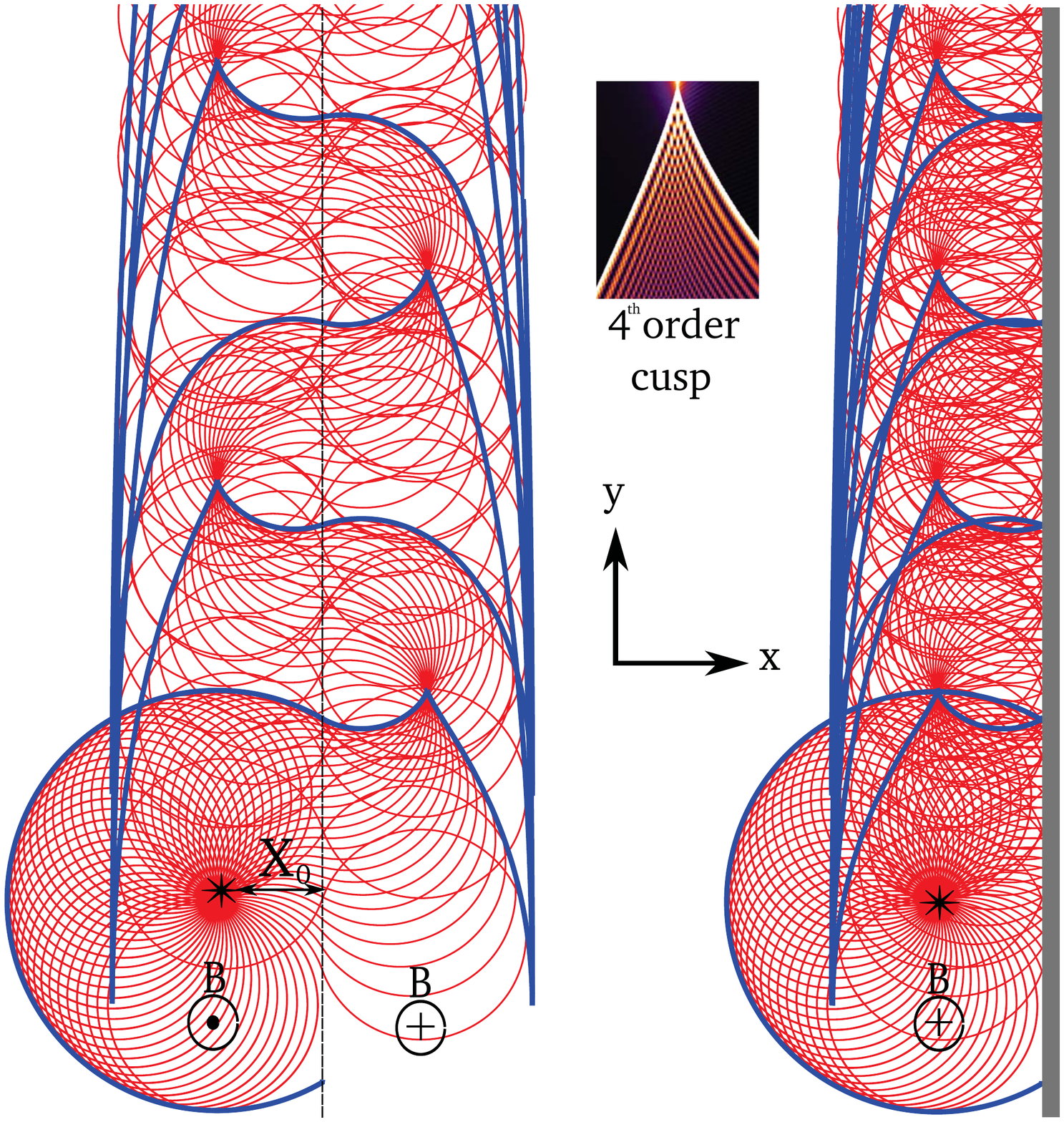}
\caption{Snake orbits (left) and skipping orbits (right) for electrons 
injected at a distance $X_0=R$ from the $\pm B$-interface/edge,
with caustics (blue) mergin at a higher-order singularity: a section of
the so called batterfly bifurcation \cite{footnote}.}
\label{Fig2}
\end{figure}

Finally, the form of caustics and cusps at the transition point between 
these two regimes, {\it i.e.}, for $X_{0}=R$, is shown in Fig. 2. 
In this case, new higher-order singularities are formed - the batterfly 
bifurcation, for which In this case higher order singularities are formed 
out of two merging pieces of caustics,
\begin{eqnarray}
u_n-u_n^c = \pm (\frac{4}{5})^{\frac{5}{4}}n^{-\frac{1}{4}} 
R^{-\frac{1}{4}}(v_n^c-v_n)^{\frac{5}{4}}.
\end{eqnarray} 
Such a singularity is characterized by two additional constraints, 
$\frac{d^N F}{d\theta^N}=0$ with $N=1,2,3,4$, and it represents a section 
of the $A_5$ butterfly bifurcation \cite{footnote}. According to the 
mathematical catastrophe theory \cite{Thom,Arnold,Berry}, such higher-order 
singularities are not stable, so that their fomation is uniquely specific 
to the infinitely sharp $\pm B$-interface. Any weak smearing of the 
interface, or an effective gauge field created for electrons by lattice 
deformations in, {\it e.g.}, the bended region of folded graphene sheet 
\cite{Guinea} replaces this transition by a precursive formation of the 
third-order singularity near the already existing cusp: 
a 'swallowtail' catastrophe \cite{footnote} consisting in the 
nucleation of a pair of cusps, followed by a gradual separation of the 
latter until the three singularities form the triplets shown in Fig. 1(b). 
All of the above results are also applicable to the electron skipping 
orbits, by folding caustics in Fig. 1 onto a single half-plane as shown 
on the r.h.s. of Fig. 2. 

\begin{figure}[h]
\includegraphics[width=0.9\columnwidth]{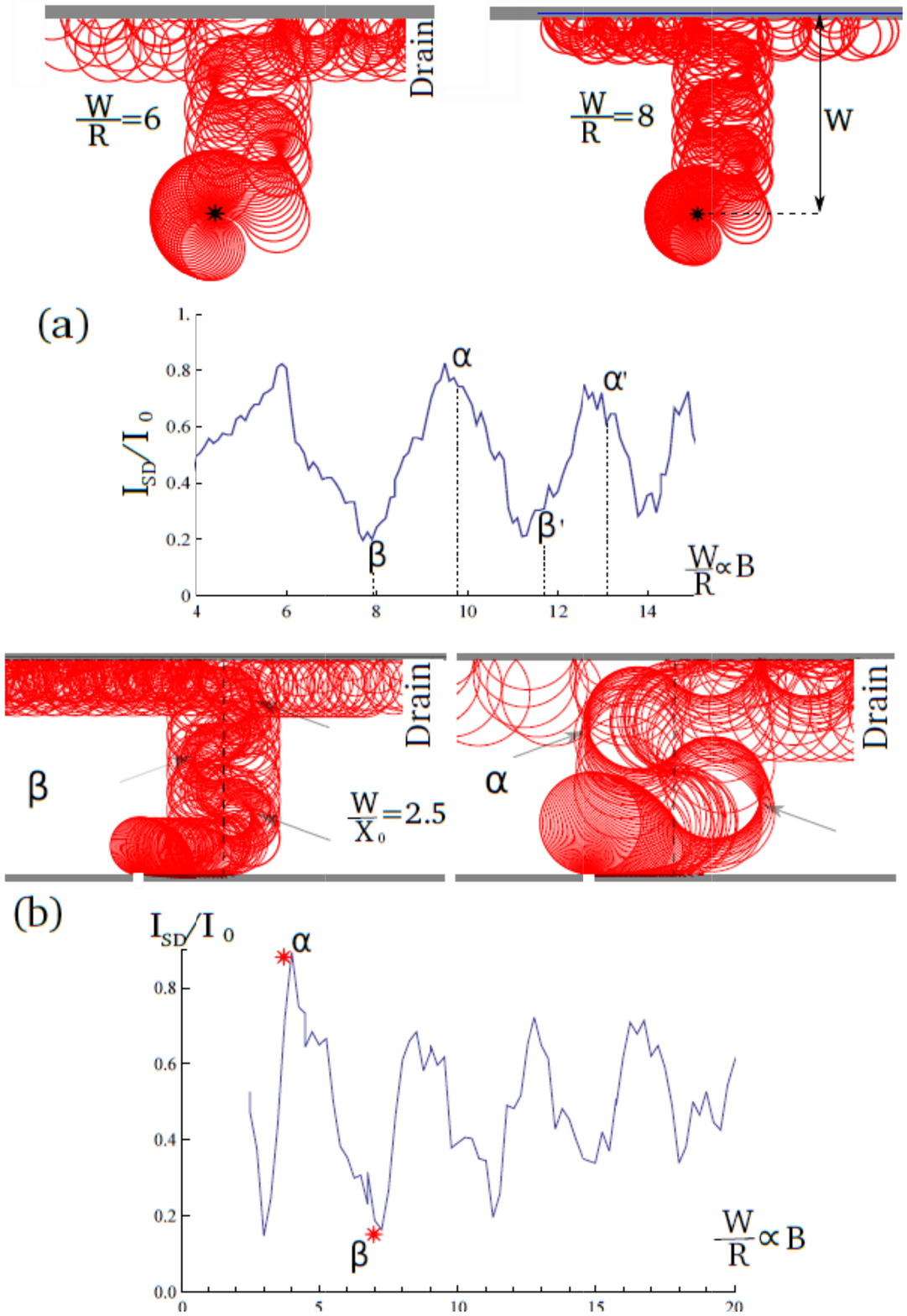}
\caption{
(a) Bunching of trajectories (for $X_0=R$) and the calculated 
magneto-oscillations of the current $I_{SD}$ in the three-terminal geometry 
(for $W=\pi ^2X_0$). Here, the injected current is registered in the 
drain placed on the right from the $\pm B$ field step. The marks 
$\alpha, \alpha '$/$\beta, \beta '$ 
relate maxima/minima of $I_{SD}$ to the cusps reaching the upper edge of 
the sample on the right/left from the field step, as marked on Fig. 1(c). 
(b) Bunching of trajectories and magneto-oscillations of the source-drain 
current in a four-terminal device with a $\pm B$-interface and width $W$. 
Two top panels illustrate families of trajectories for the conditions 
corresponding to the maximum/minimum ($\alpha$/$\beta$) of $I_{SD}$,
with positions of the cusps pointed by arrows.}
\label{Fig3}
\end{figure}

Periodic appearence of cusps of snake/skipping orbits suggests that they
can generate magneto-oscillation of non-local conductance in ballistic 
multi-terminal devices incorporating a $\pm B$-interface (or graphene fold 
in a magnetic field). Figure 3 shows the calculated magneto-oscillations 
of non-local current for two geometries of such devices. 
In Figure 3(a) we plot the magnetic-field-dependent 
proportion, $I_{SD}/I_{0}$ of the current $I_{0}$ injected from a point 
contact near the $\pm B$-interface which reaches the drain on 
the r.h.s. of the edge. This was calculated by following the propagation 
of each of the injected electrons for the time up to $10W/v_{F}$
(where $W$ is the distance from the source to the upper edge), which mimics
the effect of a finite mean free path, $\ell \sim 10W$, in the system. The
oscillations in $I_{SD}$ are the result of singularities in the ensemble of
the electron trajectories: each time when a cusp/focus on the
right from the interface reaches the upper sample edge, $I_{SD}$ experiences
a maximim, and when a singularity on the left reaches the sample edge - a
minimum. Such oscillatory behavior persist both in the regime of individual 
cusps formation and the regime of cusp triplets. However, for the lowest 
magnetic fields, such that $R/X_0 \ll 1$, the cusps in each of triplet get 
separated so much that one of those crosses the $\pm B$-interface; 
after that, the magneto-oscillations of $I_{SD}$ become rather irregular 
and loose in the amplitude. Figure 3(b) gives an example of 
magneto-oscillations of the current $I_{SD}$ in a 4-terminal device 
incoporating a $\pm B$-interface. Here, current is injected from an 
isotropic side contact at the lower edge (biased against the electrode on 
the l.h.s. at the upper edge) and registred using a drain contact placed 
at the r.h.s. at the upper edge. Similarly to Fig. 3(a), oscillations of 
$I_{SD}$ in Fig. 3(b) reflect periodic appearence (on the 
left and right hand side from the $\pm B$-interface) 
of cusps in the family of sequentially linked skipping and snake orbits. 

To summarise, we show that snake/skipping orbits of electrons injected 
at one point into in a 2D metal (at the distance $X_0$ from the 
$\pm B$-interface/edge) generically display caustic bunching and formation 
of intense local singularities - cusps, with two characteristic regimes of 
cusp formation: (i) periodic appearence of individual cusps (for $2R>X_0>R$) 
and (ii) cusp triplets (for $R>X_0$). Singularities in the distribution of 
trajectories, 
which are most intense when $R=X_0$, can lead to the magneto-oscillations in 
the non-local conductance of multi-terminal devices made, {\it e.g.}, of a 
bi-folded graphene flake. Alternatively, one can employ near-field optics 
source to generate electron-hole pairs in the heterostructure, with 
electrons placed at the energy close to the Fermi level, and, then, 
detecting the presence of singularities by measuring magnetic-field and
source-position dependences of a voltage drop between a fixed point contact
and a massive contact placed further up along the edge.

The authors thank I. Aleiner for useful discussions, Royal Society and
EPSRC for financial support, and KITPC (Beijing) for the 
hospitality during the Programme 'Spintronics and Graphene', where the 
reported study has been started.


\begin{thebibliography}{99}

\bibitem{Bohr-1911} N. Bohr. 
\textit{Dissertation}. 
Copenhagen 1911, in Niels Bohr Collected Works V.1, 276 
(Elsevier 1972).

\bibitem{Teller} E. Teller. 
Z. Phys. 67, 311 (1931).

\bibitem{Halperin} B.I. Halperin. 
Phys. Rev. B 25, 2185 (1982).

\bibitem{vanHouten} H. van Houten, B.J. van Wees, J.E. Mooij, 
C.W.J. Beenakker, J.G. Williamson, and C.T. Foxon. 
Europhys. Lett. 5, 721 (1988).

\bibitem{Beenakker} C.W.J. Beenakker, H. Van Houten, and B.J. Van Wees. 
Europhys. Lett. 7, 359 (1988).

\bibitem{Goldman} V.J. Goldman, B. Su, and J.K. Jain. 
Phys. Rev. Lett. 72, 2065 (1994).

\bibitem{Smet} J.H. Smet, D. Weiss, R.H. Blick, G. L\"{u}tjering, 
K. von Klitzing, R. Fleischmann, R. Ketzmerick, T. Geisel, G. Weimann. 
Phys. Rev. Lett. 77, 2272 (1996).

\bibitem{Westerwelt} K.E. Aidala, R.E Parrott, T. Kramer, E.J. Heller, 
R.M. Westervelt, M.P. Hanson, and A.C. Gossard.  
Nature Physics 3, 464 (2007).

\bibitem{Mints} R.G. Mints. 
JETP\ Lett. 9, 387 (1969).

\bibitem{ShekhterRozhavski} R. Shekhter and A. Rozhavski. 
Solid State Comm. 12, 603 (1973).

\bibitem{Snakes-1} J.E. M\"{u}ller. 
Phys. Rev. Lett. 68, 385 (1992).

\bibitem{Snakes-2} J. Reijniers and F.M. Peeters. 
J. Phys. Condens. Matt. 12, 9771 (2000).

\bibitem{Snakes-3} J. Reijniers, A. Matulis, K. Chang, F.M. Peeters, and
P.Vasilopoulos. 
Europhys. Lett. 59, 749 (2002).

\bibitem{magn-step-1} T. Vancura, T. Ihn, S. Broderick, K. Ensslin, 
W. Wegscheider, and M. Bichler.  
Phys. Rev. B 62, 5074 (2000).

\bibitem{magn-step-2} D.N. Lawton, A. Nogaret, S.J. Bending, D.K. Maude,
J.C. Portal, and M. Henini.  
Phys. Rev. B 64, 033312 (2001).

\bibitem{magn-step-3} M. Cerchez, S. Hugger, T. Heinzel, and N. Schulz. 
Phys. Rev. B 75, 035341 (2007).

\bibitem{Taddei} D. Rainis, F. Taddei, M. Polini, G. Le\'{o}n, F. Guinea,
and V.I. Fal'ko. 
Phys. Rev. B 83, 165403 (2011).

\bibitem{MLG} K.S. Novoselov, A.K. Geim, S.V. Morozov, D. Jiang, M.I.
Katsnelson, I.V. Grigorieva, S.V. Dubonos, and A.A. Firsov. 
Nature 438, 197 (2005).

\bibitem{GeimPNAS} K.S. Novoselov, D. Jiang, F. Schedin, T.J. Booth, 
V.V. Khotkevich, S.V. Morozov, A.K. Geim.
PNAS 102, 10451 (2005).

\bibitem{Coleman} J.N. Coleman, et al.  
Science 331, 568 (2011).

\bibitem{Thom} R. Thom. 
\textit{Structural Stability and Morphogenesis}. 
(Benjamin Press, Reading MA, 1975).

\bibitem{Arnold} V.I. Arnold. 
\textit{Singularities of caustics and wave fronts.}
Mathematics and its Applications (Springer 2002).

\bibitem{footnote} 
Singularities encountered in this work have the
following standard representation in the ADE \cite{Arnold,Orlov}
classification of catastrophes: caustics ($A_2$),
cusps ($A_3$), swallow tail ($A_4$) and
butterfly ($A_5$).

\bibitem{Orlov} Yu.A. Kravtsov,  Yu.I. Orlov. 
\textit{Caustics Catastrophes and Wave Fields.}
Springer (1999)



\bibitem{Berry} M. Berry. Adv. Phys. 25, 1 (1976).

\bibitem{ElectronicLens} J. Spector, H.L. Stormer, K.W. Baldwin, 
L.N. Pfeiffer, and K.W. West. 
Appl. Phys. Lett. 56, 1290 (1990).

\bibitem{ElectronicInverseLens} J.J. Heremans, S. von Moln\'{a}r, D.D.
Awschalom, and A.C. Gossard. 
Appl. Phys. Lett. 74, 1281 (1999).

\bibitem{Eigler} H.C. Manoharan, C.P. Lutz, and D.M. Eigler. 
Nature 403, 512 (2000).

\bibitem{Sharvin} Yu.V. Sharvin and L.M. Fisher.  
JETP Lett. 1, 152 (1965).

\bibitem{Heil} J. Heil, et al. 
Physics Reports 323, 387 (2000).

\bibitem{Weismann} A. Weismann, M. Wenderoth, S. Lounis, P. Zahn, N. Quaas,
R.G. Ulbrich, P.H. Dederichs, and S. Bl\"{u}gel. 
Science 323, 1190 (2009).

\bibitem{Guinea} M. Vozmediano, M. Katsnelson, F. Guinea.
Phys. Reports 496, 109 (2010).

\end{thebibliography}
\end{document}